\documentstyle[12pt,floats,aps]{revtex}
\newcommand{\ral}{\rangle}
\newcommand{\lal}{\langle}
\newcommand{\Img}{\mathop{\rm Im}}
\newcommand{\Real}{\mathop{\rm Re}}
\newcommand{\D}{\displaystyle}

\newcommand{\bI}{{\bf I}}
\newcommand{\cD}{{\cal D}}
\newcommand{\cH}{{\cal H}}
\newcommand{\rA}{{\rm A}}

\newcommand{\tvphi}{{\tilde{\varphi}}}
\newcommand{\vphi}{{\varphi}}
\newcommand{\tbeta}{{\tilde{\beta}}}
\newcommand{\th}{{\tilde{h}}}
\newcommand{\tH}{{\tilde{H}}}
\newcommand{\tr}{{\tilde{r}}}
\newcommand{\tR}{{\tilde{R}}}
\newcommand{\tC}{{\tilde{C}}}
\newcommand{\tcH}{{\tilde{\cH}}}
\newcommand{\hb}{{\hat{b}}}

\newcommand{\hw}{{\hat{w}}}
\newcommand{\Ox}{{\rm O}}

\newcommand{\reduction}[2]{\left.\phantom{\bigl|} #1 \right|_{#2}}
\newcommand{\C}{{\if mm {{\rm C}\mkern -15mu{\phantom{\rm t}\vrule}}
\mkern +10mu \else \leavemode \hbox{I}\kern -.17em \hbox{C} \fi}}
\newcommand{\R}{{\if mm {\rm I}\mkern -3mu{\rm R}\else \leavevmode
\hbox{I}\kern -.17em \hbox{R} \fi}}
\begin{document}
\thispagestyle{empty}
\large
\begin{center}
{\bf Bogoliubov Laboratory of Theoretical Physics\\ }
{\bf JOINT INSTITUTE FOR NUCLEAR RESEARCH\\ }
{\bf 141980 Dubna (Moscow region), Russia}
\end{center}
\vskip-3mm
\hrule{\hfill}
\vskip 2.5cm
\hfill {\large Preprint JINR E4--96--201}

\hfill {\normalsize LANL E-print {\tt nucl-th/9606012}}
\bigskip

\bigskip

\bigskip

\bigskip

{\large
\noindent  V.~B.~Belyaev$^*$, A.~K.~Motovilov$^{**}$}
\bigskip

\medskip

{\large
\noindent PERTURBATION OF EMBEDDED EIGENVALUE \\
          BY A NEAR-LYING RESONANCE}

\normalsize

\vfill

\noindent
The case of quantum-mechanical system (including electronic
molecules) is considered where Hamiltonian allows a separation,
in particular by the Faddeev method, of a weakly coupled
channel. Width (i.~e.~ the imaginary part) of the resonance
generated by a discrete spectrum eigenvalue of the separated
channel is studied in the case where main part of the
Hamiltonian gives itself another resonance. It is shown that if
real parts of these resonances coincide and, at the same time, a
coupling between the separated and main channels is sufficiently
small then the width of the resonance generated by the separated
(molecular) channel is inversely proportional to the width of
the main (nuclear) channel resonance. This phenomenon being a
kind of universal law, may play an important role increasing the
``cold fusion'' probability in electronic molecules whose
nuclear constituents have narrow pre-threshold resonances.
\bigskip

\noindent Submitted to 
{\it Teoreticheskaya i Matematicheskaya Fizika }
\vskip3cm

\hrule{\hfill}

\noindent ${}^{*}$E-mail: BELYAEV@THSUN1.JINR.DUBNA.SU

\noindent ${}^{**}$E-mail: MOTOVILV@THSUN1.JINR.DUBNA.SU

\newpage

\setcounter{page}{1}
\section{Introduction}

It is usually accepted that properties of electronic molecules
are determined exclusively by Coulombic forces acting between
nuclei and electrons and the strong interaction between nuclear
constituents plays a negligible role.  Surely, with respect to
the nuclear channels, electronic molecule is indeed observed as
a very stable object but it is evident nevertheless that any
molecular level (corresponding to the Coulomb interactions only)
looks in fact as an eigenvalue embedded into continuous spectrum
of a sub-Hamiltonian of the molecule describing its nuclear
constituents. If there are no special reasons, this means that a
coupling between the molecular and nuclear channels turn the
molecular levels into resonances%
\footnote{By resonance we understand a complex value of energy
(in an unphysical sheet) where analytic continuation of the
resolvent kernel (or its matrix element between suitable
states~\cite{ReedSimonIV}) has a pole. Such poles turn out
usually to be poles of the scattering matrix continued, too.}
(see, e.~g., Refs.~\cite{Albeverio72} --\cite{ReedSimonIV} and
references cited therein). Of course, the coupling is extremely
small due to a wide Coulombic repulsive barrier between nuclei
and short-range character of nuclear interaction. Therefore,
widths of such resonances giving a probability of a fusion of
the nuclear constituents of the molecule, are extremely small,
too.  However, we will show the situation turns out to be rather
different in the case where nuclear subsystem of a molecule has
a sufficiently narrow near-threshold (more precisely,
pre-threshold) resonance.  A number of concrete examples
concerning the nuclear systems forming near-threshold resonance
states may be extracted e.~g., from the data presented in
Ref.~\cite{ENSDF}. Among them are even such intrigue systems as
$p\,p\,{}^{16}\Ox$ and
$p\,{}^{17}\Ox$~\cite{83Ajzenberg-Selove18-20},
\cite{95Levels18-19} i.~e.  the nuclear constituents of the
usual water molecule $H_2{O}$ or hydroxyl ions $OH^-$ based on
the more rare isotope ${}^{17}\Ox$.  Possibility of influence of
near-threshold nuclear resonances on the molecular properties
was recently indicated in Ref.~\cite{BMS} where an estimation of
overlap integrals between ansatz molecular and resonance nuclear
wave functions was made for the molecules ${Li}D$ and $H_2{O}$.
There exists also a well-known example~\cite{BreunlichOthers} of
muon catalyzed fusion of deuteron and triton in the $dt\mu$
molecule where a near-threshold nuclear resonance plays a
decisive role.

Although we have in mind, first of all, the molecular systems
(which can include instead of electrons another light
negative-charged particles like muons) we deal in the paper with
more general Hamiltonians.  The only thing we use is a
possibility of a separation in them of weakly coupled
(``molecular'') channels giving initial eigenvalues
(``molecular''energies) embedded into continuous spectrum of
respective rest (``nuclear'') Hamiltonians. So that these
eigenvalues turn automatically into  ``molecular'' resonances
when a coupling between the separated ``molecular'' and rest
``nuclear'' channels is switched on.

For all the models concerned we prove the following statement.

If the ``nuclear'' channel itself has a narrow resonance with a
real part being nearly the same as the initial ``molecular''
energy then {\em the width} (i.~e. the imaginary part) {\em of
the resulting ``molecular'' resonance turns out to be inversely
proportional to the ``nuclear'' width} \/ [and increases in this
regime till its value acquires an order of the (decreasing)
``nuclear'' one].  In other words, the more narrow is the
``nuclear'' resonance, the more wide is the ``molecular'' one.
Consequently, a large increase of the decay rate of the
``molecular'' state may be observed in the case concerned, in
a contrast to the cases where such near-lying ``nuclear''
resonances with small widths are absent.

We deliberately consider in the paper certain abstract
Hamiltonians since, on the one hand, this consideration exposes
more clearly a mechanism of the effect of interplay between the
resonance widths.  On the other hand, our consideration shows
that such an enhancement of decay of a resonance state due to a
presence of another closed resonance makes a sense of an
universal law working in a wide class of quantum-mechanical
systems. The same effect has to take place also in various
problems of classical condensed matter, optics and electronics
etc., in all the cases description of which may be reduced to
spectral problems for model three-channel or two-channel
operators like~(\ref{H3}) and (\ref{H2}) or more general
Hamiltonians of the form~(\ref{Hmol}).

We start, in Section~\ref{3channel}, with an explicitly solvable
matrix model~(\ref{H3}) which demonstrates as clearly as
possible a mechanism and the effect of interplay between the
resonance widths. The model corresponds to a system of
arbitrary nature allowing to separate a main channel with
continuous spectrum and two additional weakly coupled
one-dimensional channels giving, in absence of coupling, two
discrete spectrum eigenvalues embedded into continuous spectrum
of the main channel. In the case of a molecule, the latter are
considered as trial energies for the nuclear and molecular
resonances.

If the molecular state may be considered usually as weakly
coupled with nuclear channels and, therefore, its separation
into a molecular resonance channel like channel~2 in the
Hamiltonian~(\ref{H3}) may be done comparatively easy, such a
separation of the nuclear resonance channel like channel~1 in
this Hamiltonian, may turn out to be difficult.  This is why we
specially consider in Section~\ref{Two-Channel} another model
Hamiltonian~(\ref{H2}) with explicit separation of the
``molecular'' resonance channel only.

Section~\ref{Molecule} is devoted to consideration of a model
including realistic sub-Hamiltonians for nuclear subsystems of
$N$-atomic molecules with \hbox{$N\geq 2$}.  In this case, a
separation of the molecular resonance channel is realized in
framework of an approach motivated by the Faddeev
method~\cite{Faddeev63}, \cite{MF}.

In Section~\ref{Decay} we consider a time evolution of the
molecular state in a presence again of the narrow nuclear
pre-threshold resonance.  There is shown in this section that in
a wide time interval, decay of the molecular state does have a
standard exponential character~\cite{Baz'}. The decay is obliged
first of all to the open nuclear channels and its rate is ruled
just by the inverse width of the nuclear resonance.

\section{Hamiltonian with explicit separation of both
resonance channels}\label{3channel}

First, we introduce a three-channel Hilbert space $\cH=\cH_0
\oplus \cH_1 \oplus \cH_2$, the sum of a ``main'' Hilbert space
$\cH_0$ (channel 0) and two one-dimensional spaces
$\cH_1=\cH_2=\C$ (channels 1 and 2, respectively).  Elements of
the space $\cH$ are represented as ``vectors''
$u=\left(\begin{array}{c}
u_0 \\
u_1 \\
u_2
\end{array}\right)$
with $u_0\in\cH_0$ and $u_1$, $u_2$, complex numbers, $u_1,
u_2\in\C$.  Inner product $\lal\,\cdot\,,\,\cdot\,\ral$ in $\cH$
is naturally defined via inner product
$\lal\,\cdot\,,\,\cdot\,\ral$ in $\cH_0$ as $\lal u,v\ral=\lal
u_0,v_0\ral +u_1\overline{v}_1  +u_2\overline{v}_2 $.

We deal with Hamiltonian $h$ acting in $\cH$ as a
$3\times 3$ (operator) matrix,

\begin{equation}
\label{H3}
h=\left(\begin{array}{ccc}
h_0                        & \quad b_{01}\quad    & b_{02}  \\
\lal\,\cdot\,,b_{01}\ral   &   \lambda_1          &  b_{12} \\
\lal\,\cdot\,,b_{02}\ral   & \overline{b}_{12}         & \lambda_2
\end{array}
\right)
\end{equation}
with $h_0$, a self-adjoint operator in the main Hilbert space
$\cH_0$ and $\lambda_1, \lambda_2$, the real numbers.  The
vectors $b_{01}$ and $b_{02}$ of the space $\cH_0$ realize a
coupling of the channels 1 and 2 corresponding to the initial
eigenvalues $\lambda_1$ and $\lambda_2$ with the main channel
$\cH_0$.  The complex number $b_{12}$ describes an immediate
coupling between the channels 1 and 2.

If the coupling between the channels in $h$ is
absent, $b_{01}=b_{02}=0$ and \hbox{$b_{12}=0$}, spectrum of the
Hamiltonian $h$ consists of the operator $h_0$ spectrum and two
additional discrete spectrum eigenvalues $\lambda_1$ and
$\lambda_2$ (which are interpreted as trial values
respectively, for nuclear resonance and molecule energy).  We
suppose that the Hamiltonian $h_0$ has a continuous spectrum
denoted by $\sigma_c(h_0)$, and
the eigenvalues $\lambda_1$ and $\lambda_2$
are embedded into this spectrum.  It is assumed also that
$\lambda_1, \lambda_2$ are not threshold points of the spectrum
$\sigma_c(h_0)$ and in a wide vicinity of $\lambda_1, \lambda_2$
this spectrum is absolutely continuous.  If the
coupling between the channels becomes non-trivial, the
eigenvalues $\lambda_1$ and $\lambda_2$ may go out on unphysical
sheet(s) of the energy plane turning into resonances.  The
latter are considered as poles of analytic continuation on
unphysical energy sheets of the resolvent
kernel $r(z)=(h-z)^{-1}$ or its matrix elements taken between
suitable states.

We suppose here that the resolvent $r_{0}(z)=(h_{0}-z)^{-1}$
does not have such poles in a sufficiently wide neighborhood
$\cD$ of the points $\lambda_1$ and $\lambda_2$ in
unphysical sheet(s) and the matrix elements
$\beta_{jk}(z)\equiv\lal r_0(z)b_{0j},b_{0k}\ral$, \hbox{$j,k=1,2,$}
allow analytical continuation in $z$ at least on this
neighborhood $\cD$ and
$$
c_\cD^{(j)}=\mathop{\rm inf}\limits_{z\in\cD}
|\Img\lal r_0(z)\hat{b}_{0j},\hat{b}_{0j}\ral| > 0, \quad
C_\cD=\mathop{\rm max}\limits_{j,k}\,
\mathop{\rm sup}\limits_{z\in\cD}
|\lal r_0(z)\hat{b}_{0j},\hat{b}_{0k}\ral| < \infty
$$
where $\hat{b}_{0j}=b_{0j}/\|b_{0j}\|$, $j=1,2,$ and the values
of $c_\cD^{(j)}$ and $C_\cD$ for given $\hat{b}_{0j}$
and $\hat{b}_{0k}$, depend only on $\cD$.  So that for given
(and fixed) ``structure functions'' $\hat{b}_{0j}$ one has the
following estimates:

\begin{equation}
\label{BetaEst}
c^{(j)}_\cD \|b_{0j}\|^2 \leq|\Img\beta_{jj}(z)|\leq
\tC^{(j)}_\cD \|b_{0j}\|^2
\quad\mbox{and}\quad
|\beta_{jk}(z)|\leq C_\cD \|b_{0j}\| \|b_{0k} \|
\end{equation}
where
$
\tC^{(j)}_\cD=\mathop{\rm sup}\limits_{z\in\cD}
|\Img\,\lal r_0(z)\hat{b}_{0j},\hat{b}_{0j}\ral|,\quad
$
$c^{(j)}_\cD\leq\tC^{(j)}_\cD\leq C_\cD$.

Let us consider the spectral problem $hu=zu$,
\begin{equation}
\label{H3spec}
\begin{array}{rcrcrcc}
(h_0-z) u_0 &+& b_{01} u_1 &+& b_{02} u_2 & = & 0 \\
\lal u_0,b_{01}\ral &+& (\lambda_1-z) u_1 &+&  b_{12} u_2
&=& 0\\
\lal u_0,b_{02}\ral &+& \overline{b}_{12} u_1 &+&
(\lambda_2-z) u_2 &=& 0
\end{array}
\end{equation}
for the Hamiltonian $h$. Expressing the component
$u_0=-r_0(z)b_{01} u_1 -r_0(z)b_{02} u_2$ from the first
equation~(\ref{H3spec}) and substituting it into the rest
equations, one comes immediately to an equivalent scalar system
including the components $u_1$ and $u_2$ only,

\begin{equation}
\label{H3System}
\left\{
\begin{array}{rcrc}
[\lambda_1-z-\beta_{11}(z)] u_1 & + &
[b_{12}- \beta_{12}(z)]u_2      & = 0   \\
\phantom{.}
[\overline{b}_{12}- \beta_{21}(z)] u_1     & + &
[\lambda_2-z-\beta_{22}(z)]u_2  & = 0.
\end{array}\right.
\end{equation}
Therefore, the system~(\ref{H3System}) and, hence, the
problem~(\ref{H3spec}) is solvable in $\cH$ if and only if
the energy $z$ satisfies the ``quadratic''  equation

\begin{equation}
\label{H3dispers}
(\lambda_1-z-\beta_{11})(\lambda_2-z-\beta_{22})-
(b_{12}- \beta_{12})(\overline{b}_{12}- \beta_{21})=0.
\end{equation}
It is easy to check that exactly the left part of the
equation~(\ref{H3dispers}) is present as a denominator in the
expression for the resolvent $r(z)$.  Obviously, this equation
has no solutions $z$ with $\Img z\neq 0$ in the physical sheet.
Otherwise, such $z$ were the discrete spectrum eigenvalues of
$h$. However, this is impossible since the Hamiltonian $h$ is a
self-adjoint operator.  Therefore the equation~(\ref{H3dispers})
may be solvable only in real axis and/or in unphysical sheet(s)
of the energy Riemann surface of $r_0(z)$.

At the beginning, let us consider, a special case where
$b_{12}=0$ and, say, $b_{0j}\neq 0$ but $b_{0k}=0$, $k\neq j$.
In this case, the eigenvalue $\lambda_k$ stays fixed.
At the same time the equation~(\ref{H3dispers}) turns into

\begin{equation}
\label{H2dispers}
z=\lambda_j-\lal r_0(z)b_{0j},b_{0j}\ral.
\end{equation}
If $\|b_{j0}\|$ is small enough, to prove existence of solutions
in a vicinity of \hbox{$z=\lambda_j$}, one can apply to the
equation~(\ref{H2dispers}) the fixed point
theorem.  Since, by supposition, $\lambda_j$
is not a threshold point of $\sigma_c(h_0)$, the
equation~(\ref{H2dispers}) gives us two resonances (see,
e.~g.,~\cite{KMMM-YaF88}, \cite{MotRemovalJMP})
looking in a first order of perturbation
series as

\begin{equation}
\label{H2resonance}
z^{(\pm)}_j \mathop{=}\limits_{\|b_{j0}\|\to 0}
\lambda_j-\lal r_0(\lambda_j\mp i0)b_{0j},b_{0j}\ral
+o(\|b_{j0}\|^2) ,
\end{equation}
and being situated,
respectively, in lower and upper half-planes of unphysical
sheet(s) neighboring with the physical one in a vicinity of the point
$z=\lambda_j$. The real, $E_R^{(j)}$, and imaginary,
$\Gamma_R^{(j)}/2$, parts of the resonances $z_j^{(\pm)},$
$z^{(\pm)}_j=E_R^{(j)}\pm i\D\frac{\Gamma_R^{(j)}}{2}$,
are given by
\begin{eqnarray}
\nonumber
E_R^{(j)} &=& \lambda_j-
\Real\lal r_0(\lambda_j+i0)b_{0j},b_{0j}\ral+o(\|b_{j0}\|^2), \\
\label{Gammaj}
\Gamma_R^{(j)} &=& 2\Img\lal r_0(\lambda_j+i0)b_{0j},b_{0j}\ral
+o(\|b_{j0}\|^2).
\end{eqnarray}

For analysis of the general three-channel
equation~(\ref{H3dispers}), it is convenient to resolve it  for
given $\beta_{jk}=\lal r_0(z)b_{0j},b_{0k}\ral$ as a quadratic
equation with respect to $z$:

\begin{equation}
\label{H3roots}
z=\D\frac{\lambda_1+\lambda_2-\beta_{11}-\beta_{22}}{2}\pm
\D\sqrt{
\left(\D\frac{\lambda_1-\lambda_2-\beta_{11}+\beta_{22}}{2}\right)^2
+(b_{12}-\beta_{12})(\overline{b}_{12}-\beta_{21})}
\end{equation}

We suppose the resonances generated by the both eigenvalues
$\lambda_1$ and $\lambda_2$, are narrow. This means that the
coupling of the channels in~(\ref{H3}) must be weak, i.~e.  the
previous conditions of a ``very small'' $\|b_{0j}\|$, j=1,2,
as well as the condition of a ``very small'' $|b_{12}|$ are valid.
So, the solvability of the equations~(\ref{H3roots}) in unphysical
sheet(s) may be proved again using the fixed point theorem.

Here, we consider two cases:
\medskip

1. $C_\cD \|b_{0j}\| \|b_{0k} \|\ll |\lambda_2-\lambda_1|$ for all
$j,k=1,2$ and $\|b_{12}\|\ll |\lambda_2-\lambda_1|$. One can see easily
in this case the equations~(\ref{H3roots}) give two pairs of
solutions which are practically independent on each other and
these solutions are expressed by formulae like~(\ref{H2resonance}).
\medskip

2. The case where $\lambda_1\approx\lambda_2$, but nevertheless

\begin{equation}
\label{Main}
|(b_{12}-\beta_{12})(\overline{b}_{12}-\beta_{21})|\ll
|\lambda_1-\lambda_2-\beta_{11}+\beta_{22}|^2.
\end{equation}
Namely this case is of most interest for us in the context of the
main topic of the work. We have in mind that the Hamiltonians

\begin{equation}
\label{h1h2}
h_{1}=\left(\begin{array}{cr}
h_0                         & \quad b_{01} \\
\lal\,\cdot\,, b_{01}\ral   &        \lambda_1
\end{array}
\right)
\quad\mbox{and}\quad
h_{2}=\left(\begin{array}{cr}
h_0                         & \quad b_{02}            \\
\lal\,\cdot\,, b_{02}\ral   &  \lambda_2
\end{array}
\right)
\end{equation}
are as if modelling, respectively, the nuclear subsystem of a
molecule forming a resonance state and the molecule itself in
absence of the nuclear resonance.  Leading terms of the
``nuclear'' and ``molecular'' resonances $z_1^{(\pm)}$ and
$z_2^{(\pm)}$ generated by the respective Hamiltonians $h_1$ and
$h_2$ are given by~(\ref{H2resonance}).  According
to~(\ref{BetaEst}) their widths~(\ref{Gammaj}) satisfy
inequalities
$$
c^{(1)}_\cD\|b_{01}\|^2\leq\Gamma_R^{(1)}/2
\leq\tC^{(1)}_\cD\|b_{01}\|^2
\;\mbox{ and }\;
c^{(2)}_\cD\|b_{02}\|^2\leq\Gamma_R^{(2)}/2
\leq\tC^{(2)}_\cD\|b_{02}\|^2.
$$

Since the molecular width $\Gamma_R^{(2)}$ is usually much more
small, $\Gamma_R^{(2)}\ll \Gamma_R^{(1)}$, as compared to the
nuclear one, $\Gamma_R^{(1)}$, one may require

\begin{equation}
\label{B2B1}
C_\cD\|b_{02}\|^2\ll c^{(1)}_\cD\|b_{01}\|^2.
\end{equation}
The channel $\cH_1$ belongs to the pure nuclear part of the system.
This means that, if there are no special reasons,
its constant of coupling $|b_{12}|$ with the
molecular channel $\cH_2$ must have
the same order of magnitude as $\|b_{02}\|$, i.~e.
\begin{equation}
\label{B12}
|b_{12}|\sim \|b_{02}\|\quad(\mbox{and, thereby,}
\quad C_\cD|b_{12}|^2\ll c^{(1)}_\cD\|b_{01}\|^2).
\end{equation}
From the beginning, we assume for convenience
the ``coupling constant'' $\|b_{01}\|$ to be so small that
\begin{equation}
\label{cbb}
C_\cD\|b_{01}\|<1
\end{equation}
i.~e. the ``nuclear'' resonances $z_1^{(\pm)}$
are as narrow as $\Gamma_R^{(1)}<\tC^{(1)}_\cD/C_\cD^2$.

In the condition~(\ref{B2B1}) we have
\begin{eqnarray}
\nonumber
|\lambda_1-\lambda_2-\beta_{11}+\beta_{22}|^2 &\approx&
|\lambda_1-\lambda_2-\beta_{11}|^2\geq
[\Img\beta_{11}(z)]^2\approx \\
 &\approx& [\Img\beta_{11}(\lambda_1+i0)]^2
 \approx\left(\D\frac{ \Gamma_R^{(1)} }{2}\right)^2
\geq \left(c^{(1)}_\cD\right)^2\|b_{01}\|^4.   \label{Neq1}
\end{eqnarray}

At the same time,
$$
|(b_{12}-\beta_{12})(\overline{b}_{12}-\beta_{21})|\leq
(|b_{12}|+C_\cD \|b_{01}\|\|b_{02}\|)^2.
$$
Since the conditions~(\ref{B12}) and~(\ref{cbb}) take place, we
get

\begin{equation}
\label{Neq2}
(|b_{12}|+C_\cD \|b_{01}\|\|b_{02}\|)^2\sim
\|b_{02}\|^2(1+C_\cD \|b_{01}\|)^2\leq
(c^{(2)}_\cD)^{-1}\Gamma_R^{(2)}.
\end{equation}

Let us suppose further that the ``nuclear'' resonances $z_1^{(\pm)}$
is in fact as narrow that

\begin{equation}
\label{Main0}
\Gamma_R^{(1)}\ll 1/c^{(2)}_\cD.
\end{equation}
Then, as follows from the estimates~(\ref{Neq1}) and~(\ref{Neq2}),
the condition~(\ref{Main}) is obviously satisfied if

\begin{equation}
\label{Main1}
\Gamma_R^{(2)}\ll c^{(2)}_\cD\left(\Gamma_R^{(1)}\right)^2.
\end{equation}

Now, using the expression
$\sqrt{1+\varepsilon}=1+\D\frac{\varepsilon}{2}+
O(\varepsilon^2)$ at $|\varepsilon|\ll 1$, one can rewrite the
equations~(\ref{H3roots}) for $z$ in $O(\|b_{01}\|^2)$-vicinity
of $\lambda_1$ and $\lambda_2=\lambda_1+O(\|b_{01}\|^2)$ as

\begin{equation}
\label{H3rootsfin}
\begin{array}{ccl}
z&=&\D\frac{\lambda_1+\lambda_2-\beta_{11}-\beta_{22}}{2}\pm \\
 &\pm&\D\frac{\lambda_1-\lambda_2-\beta_{11}+\beta_{22}}{2}
\left[1+\D\frac{2(b_{12}-\beta_{12})(\overline{b}_{12}-\beta_{21})}
{(\lambda_1+\lambda_2-\beta_{11}-\beta_{22})^2}+O(\varepsilon^2)
\right]
\end{array}
\end{equation}
with $\varepsilon\sim
\left(c^{(2)}_\cD\right)^{-1}
{\Gamma_R^{(2)}}/{\left(\Gamma_R^{(1)}\right)^2}.$ It
follows from~(\ref{H3rootsfin}) that in the conditions above,
the equations~(\ref{H3roots}) have solutions
$$
%
z_{\rm nucl}^{(\pm)}=
\lambda_1-\beta_{11}(\lambda_1\mp i0)+o(\|b_{01}\|^2)
+\D\frac{|b_{12}|^2}{\lambda_1-\lambda_2-\beta_{11}(\lambda_1\mp i0)}
\left[1+O(\varepsilon^2)\right]
%
$$
and

\begin{equation}
\label{ResMol}
z_{\rm mol}^{(\pm)}=
\lambda_2-\beta_{22}(\lambda_2\mp i0)+o(\|b_{02}\|^2)
-\D\frac{|b_{12}|^2}{\lambda_1-\lambda_2-\beta_{11}(\lambda_1\mp i0)}
\left[1+O(\varepsilon^2)\right]
\end{equation}

Due to~(\ref{B12}) and~(\ref{Main1})  we have
$$
\begin{array}{c}
\D\frac{|b_{12}|^2}{|\lambda_1-\lambda_2-\beta_{11}(\lambda_1\pm i0)|}
\leq 2\cdot\D\frac{|b_{12}|^2}{\Gamma_R^{(1)}}\sim
2\cdot\D\frac{\|b_{02}\|^2}{\Gamma_R^{(1)}}\leq
\D\frac{\Gamma_R^{(2)}}{c^{(2)}_\cD\Gamma_R^{(1)}}=\\
=\Gamma_R^{(1)}\cdot\D\frac{\Gamma_R^{(2)}}
{c^{(2)}_\cD\left(\Gamma_R^{(1)}\right)^2}\ll\Gamma_R^{(1)}.
\end{array}
$$
Therefore, the resonances $z_{\rm nucl}^{(\pm)}$ are only weakly
perturbed initial ``nuclear'' resonances $z_1^{(\pm)}$ with
practically the same width $\Gamma_R^{(1)}$. In a contrast to
$z_{\rm nucl}^{(\pm)}$, a difference between widths of the
resonances $z_{\rm mol}^{(\pm)}$ and $z_2^{(\pm)}$ in the case
concerned can be very large.  Such a situation takes place if
the initial ``molecular'' energy $\lambda_2$ coincides with real
part of the ``nuclear'' resonances $z_1^{(\pm)},$
$$
\lambda_2=E_R^{(1)}\approx\lambda_1-\Real\beta_{11}(\lambda_1+i0).
$$
and, in accordance with~(\ref{ResMol}),

$$
z_{\rm mol}^{(\pm)}\cong
E_R^{(1)}\pm 2i\D\frac{|b_{12}|^2}{\Gamma_R^{(1)}}.
%
$$
The width $\Gamma_R^{(m)}$ of
the resonances $z_{\rm mol}^{(\pm)}$,
$$
\Gamma_R^{(m)}=2|\Img z_{\rm mol}^{(\pm)}|\cong
4\cdot\D\frac{|b_{12}|^2}{\Gamma_R^{(1)}},
$$
being small, $\Gamma_R^{(m)}\ll\Gamma_R^{(1)}$, is nevertheless
much more large than $\Gamma_R^{(2)}\cong 2\Img
\beta_{22}(\lambda_2+i0).$ Indeed, due to~(\ref{B12})
and~(\ref{Main0}) we find

\begin{equation}
\label{RatioFinIn}
\Gamma_R^{(m)}\sim
\D\frac{\|b_{02}\|^2}{\Gamma_R^{(1)}}\sim
\Gamma_R^{(2)}\cdot\D\frac{1/c^{(2)}_\cD}{\Gamma_R^{(1)}}
\gg \Gamma_R^{(2)}.
\end{equation}
This completes the proof of our statement in the case of
three-channel Hamiltonians, concerning a crucial influence of a
narrow ``nuclear'' resonance on the width of the ``molecular''
level: {\em if the ``molecular'' energy $\lambda_2$ coincides
with the real part $E_R^{(1)}$ of the ``nuclear'' resonance then
the ``molecular'' width $\Gamma_R^{(m)}$ is {\bf inversely
proportional} to the width $\Gamma_R^{(1)}$ of this resonance
and, thereby, for a small $\Gamma_R^{(1)}$, $\Gamma_R^{(1)}\ll
1/c^{(2)}_\cD$, it can be very large compared with such a width,
$\Gamma_R^{(2)}$, in absence of the ``nuclear'' resonance}.

To get a feeling of the $c^{(2)}_\cD$ value
in a molecular case, let us suppose the Hamiltonian
$h_0$ describes indeed a few-nucleon system (constituted
by nuclei of the molecule).
Then the kernel of the resolvent component $r_0^c(z)$
corresponding to the continuous spectrum of $h_0$ allows, say, in
configuration space $\R^n$ (the dimension $n$ is determined
by number of the nucleons), the following representation~\cite{MF}
$$
r_0^c(X,X',z)=\sum\limits_\rA\int\limits_{{\R}^{n_\rA}} dp_\rA\,
\D\frac{U_\rA(X,p_\rA)\overline{U}_\rA(X',p_\rA)}
{E_\rA+p_\rA^2-z}
$$
where $X,X'$ stand for points of $\R^n$ and $\rA$, for
multi-indices (see Ref.~\cite{MF}, {\S\S}2--4  of Chapter~I)
numerating the scattering channels.  Notations $U_\rA$ are used
for respective channel wave functions (kernels of the wave
operators) and $E_\rA$, for thresholds.  The channel dimensions
$n_\rA$ are determined by numbers of clusters (nuclei) in
respective initial scattering states.

In the case, the coupling vectors $b_{02}$ have to be considered
as functions $b_{02}(X)$ belonging to a subspace corresponding to
a specific symmetry of the molecule concerned.
Calculating a jump of the kernel $r_0^c(X,X',z)$ when $z$ crosses
the real axis at the point $\lambda_2$ one finds the following
estimation for $c_\cD^{(2)}$
\begin{eqnarray*}
c_\cD^{(2)}&\sim &\Img\lal r_0^c(\lambda_2+i0)\hb_{02},\hb_{02}\ral=\\
&=&\D\sum\limits_{\rA:\,\,\, E_\rA<\lambda_2}
\pi\,[\lambda_2-E_\rA]^{(n_\rA-2)/2}\D\int\limits_{S^{n_\rA-1}}
d\hat{p}_\rA\,|\lal \hb_{02},
U_\rA(\sqrt{\lambda_2-E_\rA}\hat{p}_\rA)\ral|^2
\end{eqnarray*}
where $S^{n_\rA-1}$ stands for the unit sphere in $\R^{n_\rA}$,\,
$\hat{p}_\rA\in S^{n_\rA-1}$.

\section{Two-channel Hamiltonian with explicit separation only of
a ``molecular'' resonance channel}\label{Two-Channel}

In the present section we consider a Hamiltonian $h$
in the Hilbert space $\cH=\cH_1\oplus\C$
defined (cf.~Section~\ref{3channel}) as a matrix

\begin{equation}
\label{H2}
h=\left(\begin{array}{cr}
h_1            & \quad b \\
\lal\,\cdot\,, b\ral   &        \lambda_2
\end{array}
\right)
\end{equation}
where $h_1$ stands now for a main (``nuclear'') Hamiltonian
acting in a Hilbert space $\cH_1$ and $\lambda_2$,
$\lambda_2\in\R$, again for a trial ``molecular'' energy. A
vector $b\in\cH_1$ realizes a coupling between the channels.
Evidently, the Hamiltonian~(\ref{H3}) is a particular case
of the Hamiltonian~(\ref{H2}).

If $b\neq 0$, the eigenvalue $\lambda_2$ turns into resonances $z$
being solutions of the equation like~(\ref{H2dispers}),
\begin{equation}
\label{H2res}
z=\lambda_2-\beta(z)
\end{equation}
with $\beta(z)=\lal r_1(z)b,b\ral$, $r_1(z)=(h_1-z)^{-1}$. In a
contrast to Section~\ref{3channel} we assume here that the
``nuclear'' resonance%
\footnote{More precisely, two resonances.  Since the Hamiltonian
$h_1$ is a self-adjoint operator its resolvent obeys in the
physical sheet the symmetry condition
$[r_1(z)]^*=r_1(\overline{z})$. Then, it immediately follows
from the uniqueness principle of analytic continuation, if $z_1$
is a resonance in a sheet neighboring with the physical one then
the conjugate point $\overline{z}_1$ is also a resonance with
the same multiplicity (but maybe in another sheet).}
closed to $\lambda_2$, is present immediately in the channel 1,
i.~e.  the function $\beta(z)$ continued in an
unphysical sheet sticked in vicinity of $\lambda_2$
with the physical one, say,
along upper rim of the cut, has a pole
$z_1=E_R^{(1)}-i\D\frac{\Gamma_R^{(1)}}{2}$, with
$E_R^{(1)}\in\R$, $\Gamma_R^{(1)}>0$.
For the sake of simplicity we suppose that this
pole is simple, so that in this vicinity

\begin{equation}
\label{BetaRepres}
\beta(z)=\D\frac{a}{z-z_1}+\beta^{\rm reg}(z)
\end{equation}
with a holomorphic function $\beta^{\rm reg}(z)$.  For a fixed
``structure function'' \hbox{$\hat{b}=b/\|b\|$} we have
$|a|=C_a\|b\|^2$ with a constant $C_a>0$ determined by the
residue at $z=z_1$ of the resolvent $r_1(z)$ continued.  Note
that this residue is expressed in terms of respective resonance
wave functions, the so-called Gamow vectors, corresponding to
the resonance $z_1$ [see below formula~(\ref{Gamow})].
In fact, we suppose that this resonance corresponds to a kind of an
``almost eigenstate'' of the Hamiltonian $h_1$, i.~e. a limit
procedure is possible, in principle, with respect to a certain
parameter (or parameters) inside of $h_1$ in which
$\Gamma_R^{(1)}\to 0$ and the resonance turns into usual
eigenvalue with eigenvector $\psi_1\in\cH_1.$  In other words we
suppose that

\begin{equation}
\label{alimit}
C_a= C_a^{(0)}+o(1) \quad \mbox{as}\quad {\Gamma_R^{(1)}\to 0}
\end{equation}
and $C_a^{(0)}\equiv\lal\hat{b},
\psi_1\ral\lal\psi_1,\hat{b}\ral\neq 0$.
An example of such a situation with $C_a^{(0)}=1$
was already demonstrated in
the previous section with the Hamiltonians~(\ref{h1h2}).
The limit procedure~(\ref{alimit}) will be discussed in more detail
in Section~\ref{Perturbation}.

Analogously to $a$, we represent the regular term $\beta^{\rm
reg}(z)$ as $\beta^{\rm reg}(z)=\|b\|^2 f(z)$ with a factor
$f(z)$ depending only on $\hat{b}$. We shall suppose the vector
$\hat{b}$ to be such that, in a sufficiently wide domain $\cD$
about $\lambda_2$ concerned, in the unphysical sheet, the
function $f(z)$ is bounded, $C_\cD=\mathop{\rm
sup}\limits_{z\in\cD}|f(z)|<\infty,$ and $c_\cD=\mathop{\rm
inf}\limits_{z\in\cD}|\Img f(z)|>0$.  Therefore $|\beta^{\rm
reg}(z)|\leq C_\cD \|b\|^2$ and $c_\cD
\|b\|^2\leq|\Img\beta^{\rm reg}(z)|\leq \tC_\cD \|b\|^2$ where
$\tC_\cD=\mathop{\rm sup}\limits_{z\in\cD}|\Img f(z)|$,
$\tC_\cD\leq C_\cD$.

A substitution of the representation~(\ref{BetaRepres}) for
$\beta(z)$ into equation~(\ref{H2res}) turns the latter into
``quadratic'' equation
$$
(z-\lambda_2)(z-z_1)+a+(z-z_1)\beta^{\rm reg}(z)=0
$$
with ``solutions''

\begin{equation}
\label{H2roots}
z=\D\frac{\lambda_2+z_1-\beta^{\rm reg}}{2}\pm
\sqrt{\left(\D\frac{\lambda_2-z_1-\beta^{\rm reg}}{2}\right)^2-a}.
\end{equation}
The equations~(\ref{H2roots}) are quite analogous to the
equations~(\ref{H3roots}) being suitable for applying the fixed point
theorem when proving a solvability of~(\ref{H2res}).

We shall suppose that the coupling between channels in the
Hamiltonian~(\ref{H2}) is so weak that
\begin{equation}
\label{H2Conditions}
\left|\reduction{\beta^{\rm reg}(z)}{z\in\cD}\right|\leq
C_\cD\|b\|^2\ll\Gamma_R^{(1)}\quad\mbox{and}\quad
|a|= C_a\|b\|^2\ll\left(\Gamma_R^{(1)}\right)^2.
\end{equation}
This means

\begin{eqnarray}
\nonumber
\lefteqn{ \D\frac{|a|}{  \left|\lambda_2-z_1-
\reduction{\beta^{\rm reg}(z)}{z\in\cD}\right|^2  } =
\D\frac{|a|}{   \left|\lambda_2-E_R^{(1)}+
i \D\frac{ \Gamma_R^{(1)} }{2}-
\reduction{\beta^{\rm reg}(z)}{z\in\cD}\right|^2}\approx}\\
 &\approx & \D\frac{|a|}{\left|\lambda_2-E_R^{(1)}\right|^2+
\left(\D\frac{ \Gamma_R^{(1)} }{ 2 }\right)^2}\leq
 \D\frac{ 4|a| }{\left(\Gamma_R^{(1)}\right)^2}\leq
\D\frac{4C_a\|b\|^2}{ \left(\Gamma_R^{(1)}\right)^2 }\ll 1.
\label{NeqEst2}
\end{eqnarray}
Note that if the resonance $z_1$ was absent in the channel 0
and in such a case, one had $\beta(z)\equiv\beta^{\rm reg}(z)$,
then the eigenvalue $\lambda_2$ had generated, in the lower half-plane
$\Img z\leq 0$, the resonance
$z_2=E_R^{(2)}-i{\Gamma_R^{(2)}}/{2}$
(see Section~\ref{3channel}) with the width
$
\Gamma_R^{(2)}\approx
2\Img\beta^{\rm reg}(\lambda_2+i0)
$
satisfying inequalities
$$
   c_\cD\|b\|^2\leq\Gamma_R^{(2)}/2\leq\tC_\cD\|b\|^2.
$$
Therefore, we can compare (in terms of the
``pure nuclear'', $\Gamma_R^{(1)}$, and ``pure molecular'',
$\Gamma_R^{(2)}$, widths) the case where
a ``nuclear'' resonance, $z_1$, is present, with an opposite case
where such a resonance is absent. In particular,
the second condition~(\ref{H2Conditions}) follows
from the requirement, analogous to~(\ref{Main1}),
$$
C_a\Gamma_R^{(2)}\ll  c_\cD\left(\Gamma_R^{(1)}\right)^2.
$$

It follows from the relations~(\ref{H2roots}) considered
together with the estimates~(\ref{H2Conditions})
and~(\ref{NeqEst2}) that the domain $\cD$ includes (at $\Img z<
0$) only two roots $z_{\rm nucl}$ and $z_{\rm mol}$ of the
equation~(\ref{H2res}) with leading terms given by

\begin{eqnarray}
\label{zNucl}
 z_{\rm nucl}  & \cong & z_1+
\D\frac{a}{\lambda_2-z_1-\beta^{\rm reg}(z_1)}\cong
z_1+\D\frac{a}{ \lambda_2-z_1 },\\
\label{zMol}
z_{\rm mol} & \cong & \lambda_2-\beta^{\rm reg}(\lambda_2+i0)
-\D\frac{a}{\lambda_2-z_1-\beta^{\rm reg}(\lambda_2+i0)}\cong
\lambda_2-\D\frac{a}{\lambda_2-z_1}.
\end{eqnarray}

Due to the second condition~(\ref{H2Conditions}) we have
$\left|\D\frac{a}{\lambda_2-z_1}\right|\ll \Gamma_R^{(1)}$ and,
thereby, the resonance $z_{\rm nucl}$ represents only a very weak
perturbation of the initial ``nuclear'' resonance $z_1$.  At the
same time, a situation  with the ``molecular'' resonance is
crucially different (cf.~Section~\ref{3channel}).

In particular, if the ``molecular'' energy $\lambda_2$ coincides
with the real part $E_R^{(1)}$ of the ``nuclear'' resonance
$z_1$  then
$z_{\rm mol}=E_R^{(m)}-i\D\frac{\Gamma_R^{(m)}}{2}$ with
$E_R^{(m)}\cong\lambda_2-2\D\frac{\Img a}{\Gamma_R^{(1)}}$ and

\begin{equation}
\label{GmFin}
\Gamma_R^{(m)}\cong 4\D\frac{|\Real a|}{\Gamma_R^{(1)}},
\quad \Real a<0.
\end{equation}
Since $|\Real a|\sim C_a\|b\|^2\sim
\D\frac{C_a}{c_\cD}\Gamma_R^{(2)}$, one finds for
$\Gamma_R^{(m)}$ the estimate

\begin{equation}
\label{gmg2}
\Gamma_R^{(m)}\sim\Gamma_R^{(2)}\cdot
\D\frac{C_a/c_\cD}{\Gamma_R^{(1)}}.
\end{equation}
It should be noted in this place that, due to the
consideration~(\ref{alimit}) of the resonance $z_1$ as
an ``almost eigenvalue'', the ratio
$C_a=|a|/\|b\|^2$ is separated from zero, $C_a\geq C>0,$
as $\Gamma_R^{(1)}\to 0$. With this remark, it follows from the
estimations~(\ref{GmFin}) and~(\ref{gmg2}) as in
Section~\ref{3channel}, that, in the case of a narrow
``nuclear'' resonance  with a width
$\Gamma_R^{(1)}\ll\D{C_a}/{c_\cD}$, one has to observe a
large rise, proportional just to the factor
$\D\frac{C_a/c_\cD}{\Gamma_R^{(1)}}$
[cf.~formula~(\ref{RatioFinIn})], of the ``molecular'' width as
compared to the case where such a resonance is absent.

Concluding the section we note that if the
conditions~(\ref{H2Conditions}) do not fulfil i.~e. the coupling
between channels in the Hamiltonian~(\ref{H2}) is not small as
compared to the ``nuclear'' width $\Gamma_R^{(1)}$ then it
follows from~(\ref{H2roots}) that the molecular width
$\Gamma_R^{(m)}$ acquires itself an order of $\Gamma_R^{(1)}$.
We do not consider this case in the paper since such a situation
seems to be unreachable in electronic molecules.

\section{Consideration of a real molecule}\label{Molecule}

In this section we will prove the main statements of previous
sections but for a ``model-free'' Hamiltonian.
Namely, we consider now the Hamiltonian

\begin{equation}
\label{Hmol}
H=h_0+v_1+v_2
\end{equation}
in a Hilbert space $\cH_0$ with a self-adjoint main operator
$h_0$.  The potentials $v_1$ and $v_2$, symmetric operators, are
supposed to be such that the operator $H$ is also self-adjoint
as well as the operators $h_1=h_0+v_1$ and $h_2=h_0+v_2$ in the
same domain ${\bf D}\subset\cH_0$.

We consider such $H$ in particular as a realistic Hamiltonian
for the nuclear subsystem of a $N$-atomic molecule, $N\geq 2$.
The part $h_0$ will include in this case a sum of the kinetic
energy operator and Coulomb interactions inside of the nuclear
subsystem. The term $v_1$ will describe a strong interaction and
$v_2$, an additional effective interaction between the nuclear
constituents due to electrons. By $\cH_0$ we understand a
subspace corresponding to a specific symmetry of the molecule
(and thereby, of its nuclear subsystem, too).  This is why we
assume that continuous spectra of the operators $h_0,$ $h_1$ and
$h_2$ fill semi-infinitive intervals, respectively,
$\sigma_c(h_0)=\sigma_c(h_2)=(E_0,+\infty)$ and
\hbox{$\sigma_c(h_1)=(E_1,+\infty)$} and the lower boundary
(lower threshold) $E_1$ of the continuous spectrum
$\sigma_c(h_1)$ of $h_1$ is situated below than that, $E_0$, of
$h_0$ and $h_2$:  $\quad$ \hbox{$E_1<E_0$}.   The ``molecular''
Hamiltonian $h_2$ is supposed to have a simple isolated
eigenvalue $\lambda_2$, \hbox{$\lambda_2<E_0$}, with an
eigenfunction \hbox{$\vphi_2\in\cH_0$},
\hbox{$h_2\vphi_2=\lambda_2\vphi_2$}, \hbox{$\|\vphi_2\|=1$}.
For the sake of simplicity we suppose that all the discrete
spectrum%
\footnote{Here, in the case of a real molecule one has to take
in mind only the discrete spectrum corresponding to the subspace
$\cH_0$ above of a specific symmetry of the molecule.}
of $h_2$ consists of this point $\lambda_2$ only.  At the same
time, as in the previous sections, we assume that the (``pure
nuclear'') channel described by the Hamiltonian $h_1$, has a
narrow resonance $z_1=E_R^{(1)}-i\D\frac{\Gamma_R^{(1)}}{2}$,
$E_R^{(1)}\in\R$, $\Gamma_R^{(1)}>0$,
closed to the threshold $E_0$ so that
$E_R^{(1)}\approx\lambda_2$.

In principle, one could reduce the Hamiltonian $H$ to the matrix
models~(\ref{H3}) or~(\ref{H2}) using a standard
projection procedure. For example, to obtain from $H$ the
three-channel Hamiltonian~(\ref{H3}) one can introduce a state
$\tvphi_1\in\cH_0$, $\|\tvphi_1\|=1$, being a good approximation
of the nuclear resonance wave function $\vphi_1^{\rm res}$ [see
below formula~(\ref{Gamow})] at least at nuclear distances
so that $\lal h_1\tvphi_1,\tvphi_1\ral\approx E_R^{(1)}$. Then
one ortogonalizes the functions $\tvphi_1$ and $\vphi_2$
obtaining new orthonormal vectors $\psi_1$ and $\psi_2$ closed
respectively, to $\tvphi_1$ and $\vphi_2$. Considering the
projectors $P_1=\psi_1\lal\,\cdot\,,\psi_1\ral$,
$P_2=\psi_2\lal\,\cdot\,,\psi_2\ral$ and $P_0=I-P_1-P_2$, one
rewrites the Hamiltonian $H$ as a matrix operator, $H'=\{P_i
HP_j\}$, $i,j=0,1,2$, corresponding to the decomposition
$\cH=\cH_0\oplus\cH_1\oplus\cH_2$.  The operator $H'$ (and, of
course, the Hamiltonian $H$ itself) turns out evidently to be
unitary equivalent exactly to the operator~(\ref{H3}) with
$\lambda_j=\lal H\psi_j,\psi_j\ral$, $b_{12}=\lal
H\psi_1,\psi_2\ral$ and $b_{0j}=P_0 H\psi_j$. The $00$-component
of the matrix~(\ref{H3}) stays the same, $P_0 HP_0$, as in $H'$.
Unfortunately, the separation described of the resonance
channels is approximate and resolvent of the operator $P_0 HP_0$
has to keep poles (with rest, rather ``small'', residues) at the
molecular and nuclear resonances (corresponding to the total
Hamiltonian $H$). Account of this circumstance, in analysis of
the type~(\ref{H3dispers}) or~(\ref{H2res}) equations, requires
an additional, and cumbersome, consideration. To make an exact
(and rather simple) separation in $H$ of the nuclear and
molecular channels, we propose another natural approach
motivated by the Faddeev method~\cite{Faddeev63}, \cite{MF}.

\subsection{Separation of the molecular resonance
channel in the Faddeev approach} \label{SepFaddeev}

A study of the spectral properties of the Hamiltonians
decomposed like $H$ in a sum of a main Hamiltonian (usually the
kinetic energy operator) and few perturbations of a rather
arbitrary nature (in the example of three-body problems these
ones are usually two-body potentials) can be reduced to an
investigation of the Faddeev matrix operator (associated with
the respective Faddeev equations~\cite{Faddeev63}, \cite{MF}) on
its ``physical'' invariant subspace (see Ref.~\cite{Yakovlev}
and Refs. cited therein).

The Faddeev matrix corresponding to the decomposition~(\ref{Hmol})
looks as
\begin{equation}
\label{FadOper}
H_F=\left(\begin{array}{ccc}     h_0+v_1    &\, &  v_1 \\
                                 v_2    &\, &   h_0+v_2
\end{array}\right)=
\left(\begin{array}{ccc}     h_1     & \,&   v_1 \\
                                v_2    &\, &   h_2
\end{array}\right)
\end{equation}
being considered as operator in the Hilbert space
$\cH=\cH_1\oplus\cH_2$ with $\cH_1=\cH_2=\cH_0$.

The Hamiltonian~(\ref{FadOper}) represents a particular case of the
general $2\times 2$ matrix operators (with components $h_1$ and
$h_2$ generally independent on $v_1$ and $v_2$), some aspects of
spectral theory for which may be found in Refs.~\cite{ALMS}
and~\cite{Shkalikov}.

In the special case of $h_1=h_0+v_1$ and
$h_2=h_0+v_2$ considered, one can check that

\begin{equation}
\label{HFresol}
(H_F-z)^{-1}=(h_0-z)^{-1}\bI-(h_0-z)^{-1}
\left(\begin{array}{ccc}     v_1    &\,&   v_1 \\
                           v_2      &\,&   v_2
\end{array}\right)
(H-z)^{-1}
\end{equation}
with $\bI$, the identity operator in $\cH$.   This
allows one to conclude that although the operator
$H_F$ is non-symmetric but nevertheless it has a pure real spectrum.
Moreover, one can show (see~\cite{Yakovlev})
that the spectrum of $H_F$ consists of the (``physical'') spectrum
of $H$ and (``spurious'') spectrum of $h_0$. Eigenfunctions $\Psi$
of the initial Hamiltonian $H$ are expressed via eigenfunctions
$\psi=\left(\begin{array}{c} \psi_1 \\ \psi_2 \end{array}\right)$
of the Faddeev operator $H_F$ corresponding to
its ``physical'' spectrum as $\Psi=\psi_1+\psi_2$.

It follows also from the identity~(\ref{HFresol}) that if a
resonance appears as a pole of analytic continuation of
the resolvent kernel $(H-z)^{-1}$ than it has to be such a pole as
well for the resolvent $(H_F-z)^{-1}$.
An inverse statement is also valid if
the resonance is not a pole of respective continuation of the
resolvent kernel $(h_0-z)^{-1}$.

Let $Q_2$ be the projector on a subspace of $\cH_2$ orthogonal
to the ``molecular'' eigenfunction $\vphi_2$,
$Q_2=I-\vphi_2\lal\,\cdot\,,\vphi_2\ral$.  A separation of
contributions to $h_2$ from the eigenvalue $\lambda_2$ and the
rest spectrum allows us to rewrite the
Hamiltonian~(\ref{FadOper}) in the form very closed to the
model~(\ref{H2}). Namely, we rewrite this Hamiltonian as the
matrix operator

\begin{equation}
\label{htF}
H'_F=\left(\begin{array}{ccc}
   \tH_F                     &\,\,&  {\bf b}_{12} \\
   \lal\,\cdot\,, {\bf b}_{21}\ral  &\,\,& \lambda_2
\end{array}\right)
\end{equation}
considered in the Hilbert space $\tcH\oplus\C$ where
$\tcH=\cH_1\oplus Q_2\cH_2$.
By $\tH_F$ we understand the operator in $\tcH$ defined as
$$
\tH_F=\left(\begin{array}{ccc}
   h_1   &\,\,& v_1 Q_2 \\
   Q_2 v_2  &\,\,& \th_2
\end{array}\right)
$$
where $\th_2$ stands for a part of the Hamiltonian $h_2$ in
$Q_2\cH_2$.
The vectors $ {\bf b}_{12}$ and  $ {\bf b}_{21}$ of the space
$\tcH$ determining a coupling between the channels
in~(\ref{htF}), look as $ {\bf b}_{12}=\left(\begin{array}{c} b_{12}
\\ 0 \end{array}\right)$ with $b_{12}=v_1\vphi_2$,
$b_{12}\in\cH_1$,
and
$ {\bf b}_{21}=\left(\begin{array}{c} b_{21} \\ 0 \end{array}\right)$
with $b_{21}=v_2\vphi_2$, $b_{21}\in\cH_1$.

The resolvent $(H'_F-z)^{-1}$ of the reduced Faddeev operator
$H'_F$ may be written as

\begin{equation}
\label{FadReduced}
(H'_F-z)^{-1}=\left(\begin{array}{ccc}
\tR_F(z)+\D\frac{\tR_F(z) {\bf b}_{12}
\lal\tR_F(z)\,\cdot\,, {\bf b}_{21}\ral}
{D(z)}      &\,&
-\D\frac{\tR_F(z) {\bf b}_{12}}{D(z)}\\
-\D\frac{\lal\tR_F(z)\,\cdot\,,
{\bf b}_{21}\ral}{D(z)} &&\D\frac{1}{D(z)}
\end{array}\right)
\end{equation}
where $\tR_F(z)$ stands for the resolvent of the Hamiltonian
$\tH_F$, {$\tR_F(z)=(\tH_F-z)^{-1}$},
and $D(z)$, for the scalar function

\begin{equation}
\label{D}
D(z)=\lambda_2-z-\tbeta(z)
\end{equation}
with

\begin{equation}
\label{gth1}
\tbeta(z)=\lal\tR_F(z) {\bf b}_{12}, {\bf b}_{21}\ral\equiv
\lal g_1(z)b_{12},b_{21}\ral.
\end{equation}
Here, $g_1(z)$ is a generalized resolvent of the operator $\tH_F$,

\begin{equation}
\label{g1}
g_1(z)=\left(h_1-z-v_1 Q_2\tilde{r}_2(z)Q_2 v_2\right)^{-1}
\end{equation}
where $\tr_2(z)=(\th_2-z)^{-1}$.

Since the expression~(\ref{htF}) represents an unitary
equivalent form of the Faddeev operator $H_F$, singularities of
the resolvent $(H'_F-z)^{-1}$ in the physical sheet have to
coincide with those of $(H_F-z)^{-1}$, being located in the
spectra of the Hamiltonians $h_0$ and/or $H$, i.~e. in the real
axis only. This is an advantage of the
representation~(\ref{FadReduced}), in a contrast to the
formula~(\ref{HFresol}), to manifest explicitly that a
perturbation of the molecular eigenvalue $\lambda_2$, in a
presence of the nuclear channels, may be computed solving the
equation $D(z)=0$, or equivalent, the equation

\begin{equation}
\label{FadApRes}
z=\lambda_2-\tbeta(z).
\end{equation}
According to the formulae~(\ref{FadReduced}) and~(\ref{D}), the
roots of the latter turn out automatically to be poles of the
resolvent $(H_F-z)^{-1}$. Hence, such roots in the physical
sheet may exist in the real axis only. Searching for the roots
of the equation continued in unphysical sheet(s), one can find,
among them, resonance(s) engendered by the initial molecular
eigenvalue $\lambda_2$.

\subsection{Perturbation of the nuclear resonance due to
a continuous spectrum component of the molecular Hamiltonian}
\label{Perturbation}

Evidently, an analysis of the equation~(\ref{FadApRes}) in the
case of ``small'' coupling vectors $b_{12}$ and/or $b_{21}$ may
be carried out in the same way as the analysis of the analogous
equation~(\ref{H2res}) in the previous section.  And, if there
exists a pole of $\tbeta(z)$ having a real part closed to
$\lambda_2$, the effect of the inversely proportional growth of
the ``molecular'' width as the ``nuclear'' width decreases, will
appear again.  We have a candidate for such a pole,
the initial ``pure nuclear'' resonance $z_1$ which is, of
course, perturbed due  to the presence in $\tH_F$ of the extra
channel with $\th_2$.  Now we will formulate certain sufficient
conditions when such a perturbation of $z_1$ is small as
compared to $\Gamma_R^{(1)}$. It should be noted that
in real molecular systems such a property is very natural
from physical reasons.

First we remind that the eigenvalue $\lambda_2$ is situated
below the threshold $E_0$. Therefore, we consider analytic
continuation of the resolvent kernel(s) $(H'_F-z)^{-1}$ and,
thereby, the form $\tbeta(z)$ on unphysical sheet%
\footnote{In definitions of Ref.~\cite{MotovilovRes}
for a three-body problem, this one
is called a {\em two-body} sheet.}
neighboring below $E_0$ with the physical one in a vicinity of
the point $\lambda_2$.  The energy $E_0$ is, at the same time, a
lower boundary of the continuous spectrum of $h_2$ and, thereby,
of $\th_2$. This means that the resolvent $\tilde{r}_2(z)$,
included in~(\ref{g1}),  will stay during such a continuation in
the physical sheet. Of course a domain $\cD$ about $\lambda_2$
where the continuation is provided in the unphysical sheet
concerned, will be chosen to be ``sufficiently small'', with a
diameter of order of $\Gamma_R^{(1)}$ and, to be separated from
the ray $[E_0,+\infty)$.

Let us suppose that the analytic continuation of the resolvent
kernel $r_1(z)$ in a vicinity of the resonance pole $z_1$
admits the representation with explicitly factorized residue,
\begin{equation}
\label{Gamow}
r_1(z)=\D\frac{A\vphi_1^{\rm res}\lal\,\cdot\,,
\tilde{\vphi}_1^{\rm res}\ral}{z_1-z}+\tilde{r}_1(z).
\end{equation}
Here, the (generalized) eigenvectors (the so-called Gamow
states) $\vphi_1^{\rm res}$ and $\tilde{\vphi}_1^{\rm res}$ of
the Hamiltonian $h_1$ are assumed to be specific (including in
the coordinate representation only outgoing waves with
exponentially increasing asymptotics) solutions of the
Schr\"odinger equation $h_1\psi_1=z\psi_1$, respectively at
$z=z_1$ and $z=\overline{z}_1$. A complex number $A$ is a
``normalization'' constant. The term $\tilde{r}_1(z)$ represents
a regular at $z=z_1$ summand of $r_1$.  Surely, as the resolvent
kernel $r_1(z)$ itself, the terms of the
representation~(\ref{Gamow}) have to be understood in the
unphysical sheet in a sense of the generalized functions
(distributions) like e.~g., in Ref.~\cite{MotovilovRes}. This
means in particular that only sufficiently ``good'' elements
(e.~g.,  sufficiently quickly decreasing test functions in the
coordinate representation) forming a dense subset $\cH_1^{(0)}$
in $\cH_1$, can be substituted instead of ``$\,\cdot\,$'' into
the products $\lal\,\cdot\,, \vphi_1^{\rm res}\ral$ and
$\lal\,\cdot\,, \tilde{\vphi}_1^{\rm res}\ral$.

The near-resonance representation of the Green
function~(\ref{Gamow}) is approved in the two-body problems
e.~g., in framework of the Jost formalism~\cite{OnJost}. A
possibility of such a representation in the three-body problems
follows from the results of Ref.~\cite{MotovilovRes}, at least
in the case of the quickly decreasing, in the $x$-space,
interactions and thereby, in the cases where one can cut the
Coulombic potentials between nuclear constituents, say, at
molecular distances.

As in Section~\ref{Two-Channel} we assume that the resonance
$z_1$ is a kind of a discrete spectrum eigenvalue going from the
real axis. So that, a limit procedure is as if possible where
$\Gamma_R^{(1)}\to 0$ and the Gamow states $\vphi_1^{\rm res}$
and $\tilde{\vphi}_1^{\rm res}$ turn (in the weak sense) into a
``normal'' eigenfunction $\vphi_1$ (the same for both of them),
i.~e. for any $f\in\cH_1^{(0)}$

\begin{equation}
\label{Limit}
\lal\vphi_1^{\rm res},f\ral
\longrightarrow
\lal\vphi_1,f\ral, \quad
\lal\tilde{\vphi}_1^{\rm res},f\ral
\longrightarrow
\lal\vphi_1,f\ral \quad\mbox{\rm and}\quad
A\longrightarrow 1/\|\vphi_1\|^2
\end{equation}
as $\Gamma_R^{(1)}\to 0$.

Let the strong, $v_1$ and molecular, $v_2$, potentials  be such that
the product
$$
w_{12}(z)\equiv v_1 Q_2\tilde{r}_2(z)Q_2 v_2
$$
is an uniformly bounded operator-valued function of $z\in\cD$ with
a finite and rather small norm
$$
    \|w_{12}\|_\cD=\mathop{\rm sup}\limits_{z\in\cD}
   \|v_1 Q_2\tilde{r}_2(z)Q_2 v_2\|.
$$
At the same time, let the width $\Gamma_R^{(1)}$ satisfy the
condition

\begin{equation}
\label{GammaSmall}
\Gamma_R^{(1)}\ll E_R^{(1)}- E'_{1}
\end{equation}
where $E'_{1}$ is the nearest from below (may be, differing from
$E_1$) threshold of the continuous spectrum of $h_1$  (just such
a situation is expected to take place in a real molecule).  Then
one can assume that a kernel of the normalized product
$\hw_{12}(z)=\D\frac{w_{12}(z)}{\|w_{12}\|}$ falls off in
coordinate space as quickly that it may be applied, in
condition~(\ref{GammaSmall}), to $\vphi_1^{\rm res}$,
$\tilde{\vphi}_1^{\rm res}$ and $\tilde{r}_1(z)$, so that, in
particular, $\hw_{12}(z)\vphi_1^{\rm res}\in\cH_1^{(0)}$ and
$\hw^{*}_{12}(z)\tilde{\vphi}_1^{\rm res}\in\cH_1^{(0)}$. One can
assume, moreover, that the matrix element
$\lal\hw_{12}(z)\vphi_1^{\rm res},\tilde{\vphi}_1^{\rm res}\ral$
makes a sense for $z\in\cD$ as well as the cross products
$\lal\hw_{12}(z)\vphi_1^{\rm res},\tilde{r}_1(z)\,\cdot\,\ral$,
$\lal\hw_{12}(z)\tilde{r}_1(z)\,\cdot\,, \tilde{\vphi}_1^{\rm
res}\ral$ and $\lal\hw_{12}(z)\tilde{r}_1(z)\,\cdot\,,
\tilde{r}_1(z)\,\cdot\,\ral$ where one has to substitute, instead
of ``$\,\cdot\,$'', the elements of a dense subset $\cH_1^{(0)}$
mentioned above.  We suppose further that the value of
$\|w_{12}\|_\cD$ is as small that, for a fixed
$\hat{w}_{12}(z)$, the following estimate takes place

\begin{equation}
\label{v1v2norm}
\mathop{\rm sup}\limits_{z\in\cD}
|A\lal w_{12}(z)\vphi_1^{\rm res},\tilde{\vphi}_1^{\rm res}\ral|\leq
\|w_{12}\|_\cD\,\cdot\,\mathop{\rm sup}\limits_{z\in\cD}
|A\lal \hat{w}_{12}(z)\vphi_1^{\rm res},\tilde{\vphi}_1^{\rm res}\ral|
\ll \Gamma_R^{(1)}.
\end{equation}

With these presuppositions one can prove already that a shift of
the initial ``nuclear'' resonance $z_1$ due to a presence of the
$\th_2$ channel in $\tH_F$ is indeed very small as compared to
the width $\Gamma_R^{(1)}$.  To show this, we consider the
Lippmann-Schwinger equation for the generalized resolvent
$g_1(z)$,
$$
g_1(z)=r_1(z)-r_1(z)w_{12}(z)g_1(z).
$$
In the conditions above this equation may be continued in the
domain $\cD$ of the unphysical sheet concerned.  To give an
``explicit'' representation for $g_1(z)$, we shall use a
solution $\tilde{r}'_1(z)$ of another Lippmman-Schwinger
equation in the same unphysical sheet,

\begin{equation}
\label{r1prime}
\tilde{r}'_1(z) =\tilde{r}_1(z)-
\tilde{r}_1(z) w_{12}(z)\tilde{r}'_1(z).
\end{equation}
With a fixed $\hat{w}_{12}(z)$ and sufficiently small
$\|w_{12}\|_\cD$, the unique solvability of the
last equation may be easily proved for $z\in\cD$.
One can show in fact that the solution of~(\ref{r1prime}) belongs
to a class of kernels satisfying the same conditions concerning
the cross products with $w_{12}(z)$ as $\tilde{r}_1(z)$.

Then, representing the resolvent $r_1(z)$ in the
form~(\ref{Gamow}) one finds immediately that
the generalized resolvent $g_1(z)$ looks as

\begin{equation}
\label{g1final}
g_1(z)=A\D\frac{[I-\tr'_1(z)w_{12}(z)]\vphi_1^{\rm res}
\lal\,\cdot\,,\tilde{\vphi}_1^{\rm res}\ral[I-w_{12}(z)\tr'_1(z)]}
{F_1(z)}+\tr'_1(z)
\end{equation}
with

\begin{equation}
\label{Fres}
F_1(z)=z_1-z+
A\lal w_{12}(z)\vphi_1^{\rm res},\tilde{\vphi}_1^{\rm res}\ral
+A\lal w_{12}(z)\tilde{r}'_1(z) w_{12}(z)
\vphi_1^{\rm res},\tilde{\vphi}_1^{\rm res}\ral
\end{equation}

It follows from the expression~(\ref{g1final}) that
the initial nuclear resonance $z_1$ transforms
into a solution $z$ of the equation

\begin{equation}
\label{tz1}
F_1(z)=0.
\end{equation}
Thus, due to~(\ref{Fres}), the resonance
$\tilde{z}_1$ generated, instead of $h_1$,
by the Hamiltonian $\tH_F$, has the
following asymptotics (as $\|w_{12}\|_\cD\to 0$):
$$
\tilde{z}_1=
z_1+A\lal w_{12}(z_1)\vphi_1^{\rm res},\tilde{\vphi}_1^{\rm res}\ral
+O(\|w_{12}\|_\cD^2),
$$
and, thereby, according to the supposition~(\ref{v1v2norm})
the perturbation of $z_1$ due to $\th_2$ is small indeed,
$$
|\tilde{z}_1-z_1|\ll \Gamma_R^{(1)}.
$$

\subsection{Interplay between molecular and nuclear widths}

Let us return now to the equation~(\ref{FadApRes}).
Additional assumptions are accepted here that the nuclear,
$v_1$, and molecular, $v_2$, potentials and the ``molecular''
eigenfunction $\vphi_2$ are such that the products
$\lal b_{12},\tilde{\vphi}_1^{\rm res}\ral$ and
$\lal {\vphi}_1^{\rm res},b_{21}\ral$ make a sense
(i.~e. the both functions $b_{12}$ and $b_{21}$ are elements
of the dense subset $\cH_1^{(0)}$ above)
as well as the products
$\lal w_{12}(z)\tr'_1(z)b_{12},\tilde{\vphi}_1^{\rm res}\ral$
and
$\lal \tr'_1(z)w_{12}(z)\vphi_1^{\rm res},b_{21}\ral$,\,
$z\in\cD$. Then one can obviously rewrite the function
$\tbeta(z)$ for $z\in\cD$
exactly as~(\ref{H2res}),
$$
\tbeta(z)=\D\frac{\tilde{a}}{z-\tilde{z}_1}+
\tbeta^{\rm reg}(z).
$$
Here, $\tilde{a}$ stands for a residue of this function at
the pole $z=\tilde{z}_1$
and $\tilde{\beta}^{\rm reg}(z)$, for a regular part.
Therefore, the equation~(\ref{FadApRes}) may be rewritten
as~(\ref{H2roots}), i.~e.

\begin{equation}
\label{HFroots}
z=\D\frac{\lambda_2+\tilde{z}_1-\tbeta^{\rm reg}}{2}\pm
\sqrt{\left(\D\frac{\lambda_2-\tilde{z}_1-
\tbeta^{\rm reg}}{2}\right)^2-\tilde{a}}.
\end{equation}
It follows from the representation~(\ref{g1final}) that
for a given (and fixed) ``structure functions'' $\hat{w}_{12}(z)$,
$\hat{b}_{12}=\D\frac{b_{12}}{\|b_{12}\|}$ and
$\hat{b}_{21}=\D\frac{b_{21}}{\|b_{21}\|}$
the following asymptotical formulae keep true for
$\tilde{a}$ and $\tbeta^{\rm reg}$ as $\|w_{12}\|_\cD\to 0$:
\begin{equation}
\label{EstABeta}
\tilde{a}=a+O(\|w_{12}\|_\cD\|b_{12}\|\|b_{21}\|), \quad
\tbeta^{\rm reg}(z)=\beta^{\rm reg}(z)
+O(\|w_{12}\|_\cD\|b_{12}\|\|b_{21}\|)
\end{equation}
where
$a=-A\lal b_{12},\tilde{\vphi}_1^{\rm res}\ral
\lal {\vphi}_1^{\rm res},b_{21}\ral$
and $\beta^{\rm reg}(z)=\lal \tr_1(z)b_{12},b_{21}\ral.$

We suppose that at least the function $b_{12}=v_1\vphi_2$ is
so small%
\footnote{Due to internuclear Coulombic barrier suppressing the
molecular eigenfunction at nuclear distances}
that
$$
|\tbeta^{\rm reg}(z)|\approx|\lal\tr_1(z)b_{12},b_{21}\ral|
\leq C_\cD \cdot\|b_{12}\|\cdot\|b_{21}\|\ll\Gamma_R^{(1)},
$$
\begin{equation}
\label{ta}
|\tilde{a}|\approx|a|=C_a\cdot
\|b_{12}\|\cdot\|b_{21}\|\ll\left(\Gamma_R^{(1)}\right)^2.
\end{equation}
Notations $C_\cD$ and $C_a$ make now the following sense:
$$
C_\cD=\mathop{\rm sup}\limits_{z\in\cD}
 |\lal\tr_1(z)\hb_{12},\hb_{21}|, \quad
C_a=|A|\cdot|\lal\hb_{12},\tilde{\vphi}^{\rm res}_1\ral
\lal\vphi_1^{\rm res},\hb_{21}\ral|.
$$

Repeating almost literally a discussion of the
equations~(\ref{H2roots}) in Section~\ref{Two-Channel} we show
that the equations~(\ref{HFroots}) have only two solutions in
the domain $\cD$,\, $z_{\rm nucl}$ in the case of sign ``$-$''
and $z_{\rm mol}$ in the case of sign ``$+$'', leading terms of
which are given by formulae~(\ref{zNucl}) and~(\ref{zMol}).

If the molecular energy $\lambda_2$ ``almost'' coincides with
the real part $E_R^{(1)}$ of the nuclear resonance $z_1$, \,
$|E_R^{(1)}-\lambda_2|\ll\Gamma_R^{(1)}$, one gets

\begin{eqnarray}
\nonumber
z_{\rm nucl} &\cong& z_1-i\D\frac{2a}{\Gamma_R^{(1)}}, \\
\label{FzMol}
z_{\rm mol} &\cong& \lambda_2+i\D\frac{2a}{\Gamma_R^{(1)}}.
\end{eqnarray}
It is evident that the pole $z_{\rm nucl}$, being a weak [since,
in view of~(\ref{ta}),
\hbox{$|a|/\Gamma_R^{(1)}\ll\Gamma_R^{(1)}$}] perturbation of
the resonance $z_1$, belongs to the unphysical sheet concerned.
Analogous assertion for the molecular resonance $z_{\rm mol}$
follows, in principle, from the statement above that the
equation~(\ref{FadApRes}) can have at $\Img z>0$ no solutions in
the physical sheet. However, at least for small
$\Gamma_R^{(1)}$, one can immediately check the inequality
$\Real a <0$ holds true guaranteeing a location of the resonance
$z_{\rm mol}$ in the unphysical sheet.  Indeed, we note that

\begin{eqnarray}
\nonumber
\lal\vphi_1^{\rm res},v_2\vphi_2\ral &=&
-\lal\vphi_1^{\rm res},(h_0-\lambda_2)\vphi_2\ral =\\
\nonumber
 &=&-\lal(h_0-z_1)\vphi_1^{\rm res},\vphi_2\ral+(\lambda_2-z_1)
\lal\vphi_1^{\rm res},\vphi_2\ral=\\
\nonumber
&=& \lal v_1\vphi_1^{\rm res},\vphi_2\ral +
(\lambda_2-z_1)\lal v_1\vphi_1^{\rm res},\vphi_2\ral.
\end{eqnarray}
Therefore,
$$
a=-A\lal v_1\vphi_2,\tilde{\vphi}_1^{\rm res}\ral
\lal \vphi_1^{\rm res},v_1\vphi_2\ral+
A(z_1-\lambda_2)
\lal\vphi_2,v_1\tilde{\vphi}_1^{\rm res}\ral
\lal\vphi_1^{\rm res},\vphi_2\ral.
$$
Remind, we suppose the nuclear resonance $z_1$ to be such
that a limit procedure~(\ref{Limit}) is possible.
This means

\begin{equation}
\label{aLimit}
a\mathop{=}\limits_{\Gamma_R^{(1)}\to 0}
-\D\frac{|\lal\vphi_1,v_1\vphi_2\ral|^2}
{\|\vphi_1\|^2}+o(1)+i\D\frac{\Gamma_R^{(1)}}{2}
\left(\D\frac{\lal v_1\vphi_2,\vphi_1\ral
\lal\vphi_1,\vphi_2\ral}
{\|\vphi_1\|^2}+o(1)\right)
\end{equation}
So that
$$
\Real a=-\D\frac{|\lal \vphi_1,v_1\vphi_2\ral|^2}
{\|\vphi_1\|^2}+o(1)
$$
is really negative, at least for sufficiently small
widths $\Gamma_R^{(1)}$. (Of course, we exclude here the cases
of an incidental degeneration where
$\lal\vphi_1,v_1\vphi_2\ral=0$.)
So that, according to~(\ref{FzMol}),
the width $\Gamma_R^{(m)}$ of the resonance
$z_{\rm mol}$ depends on $\Gamma_R^{(1)}$ as $\Gamma_R^{(1)}\to 0$
exactly as~(\ref{GmFin}),
$\Gamma_R^{(m)}\cong 4\D\frac{|\Real a|}{\Gamma_R^{(1)}}.$

To complete discussion of the equation~(\ref{FadApRes}), let us
compare the behavior of $\Gamma_R^{(m)}$ in the presence of the
nuclear near-threshold resonance $z_1$ with an opposite case
where such a resonance is absent and, thus, the function
$\tbeta(z)$, $\tbeta(z)\equiv\tbeta^{\rm reg}(z)$, is
holomorphic in $\cD$.  In this case, the
equation~(\ref{FadApRes}) has in $\cD$ only one solution
$z_R^{(2)}$, leading terms of which are given with account of
second formula~(\ref{EstABeta}) by
$$
z_R^{(2)}\cong\lambda_2-\lal r_1(\lambda_2+i0)b_{12},b_{21}\ral=
\lambda_2-\lal r_1(\lambda_2+i0)v_1 \vphi_2,v_2 \vphi_2\ral.
$$
Since $v_2\vphi_2=-(h_2-\lambda_2)\vphi_2$, one finds
$$
z_R^{(2)}\cong\lambda_2+\lal v_1\vphi_2,\vphi_2\ral-
\lal r_1(\lambda_2+i0)v_1 \vphi_2,v_1 \vphi_2\ral.
$$
Therefore, the width $\Gamma_R^{(2)}$ of the resonance $z_R^{(2)}$
is given by
$$
\Gamma_R^{(2)}\cong 2\Img
\lal r_1(\lambda_2+i0)v_1 \vphi_2,v_1 \vphi_2\ral.
$$
This means that one has, in presence of the nuclear resonance $z_1$,
to compare the width $\Gamma_R^{(m)}$ with the value
$
|\Img\tbeta^{\rm reg}(\lambda_2+i0)|\approx
|\Img\lal \tilde{r}_1(\lambda_2+i0)v_1 \vphi_2,v_1 \vphi_2\ral|
\geq c'_\cD\|b_{12}\|^2
$
where
$
c'_\cD=\mathop{\rm inf}\limits_{z\in\cD}
|\lal\tr_1(z)\hb_{12},\hb_{12}\ral|.
$
As a result we come to (a rough) estimation like~(\ref{gmg2}),

\begin{equation}
\label{Faddeevgmg2}
\Gamma_R^{(m)}\sim\Gamma_R^{(2)}\cdot
\D\frac{C'_a/c'_\cD}{\Gamma_R^{(1)}}.
\end{equation}
with
$$
C'_a=\D\frac{1}{\|b_{12}\|^2}\,
|A\lal v_1\vphi_2,\tvphi_1^{\rm res}\ral
\lal\vphi_1^{\rm res},v_1\vphi_2\ral|
%
%
%
\mathop{\cong}\limits_{\Gamma_R^{(1)}\to 0}
\D\frac{|\lal\hb_{12},\vphi_1\ral|^2}{\|\vphi_1\|^2}.
$$
Again, as in Sections~\ref{3channel} and~\ref{Two-Channel} we
find that if the nuclear resonance is narrow, $\Gamma_R^{(1)}\ll
C'_a/c'_\cD$ then a large increasing of the molecular width,
proportional just to the factor
$\D\frac{C'_a/c'_\cD}{\Gamma_R^{(1)}}$, has to be observed. In
other words, the nuclear fusion reaction in the molecule
concerned is considerably enhanced.

Note in a conclusion that the formulae~(\ref{FzMol})
and~(\ref{aLimit}) may be used for a practical estimation of the
width $\Gamma^{(m)}_R$.  Here, a crucial role belongs to the
values ``$\vphi(0)$'' (cf.~the Jackson formula~\cite{Jackson})
of the molecular function $\vphi_2$ at small (nuclear) distances
where the strong interaction $v_1$ is localized in configuration
space.  Surely, such estimations for concrete molecules require
numerical computations.

\section{Exponential decay of the molecular state}
\label{Decay}

In Section~\ref{Molecule}, we have accepted a description of the
nuclear subsystem of a molecule by a (realistic) effective
Hamiltonian~(\ref{Hmol}).  Let us suppose now that an initial
state of the molecule corresponds exactly to the pure
``molecular'' wave function $\vphi_2$, \,
$(h_2-\lambda_2)\vphi_2=0$, \, $\|\vphi_2\|=1$.  Then, a
consequent evolution (in time $t$) of the nuclear subsystem is
described by a solution $\Psi(t)$ of the Cauchy problem
$$
    i\D\frac{d\Psi}{dt}=H\Psi, \quad \reduction{\Psi}{t=0}=\vphi_2.
$$
Probability $P_{\rm mol}(t)$ at a time moment $t$ to find the
subsystem still in the molecular state $\vphi_2$ is given by
$$
P_{\rm mol}(t)=|\lal\Psi(t),\vphi_2\ral|^2.
$$
Since $\lambda_2<E_0$ and, thereby, the continuous spectrum
channel of the ``molecular'' Hamiltonian $h_2$ is closed down,
the remainder $1-P_{\rm mol}(t)$ determines a decay probability
of the state $\vphi_2$ into open channels (branches) of
continuous spectrum of the nuclear Hamiltonian $h_1$. The latter
correspond to all possible variants of synthesis of the nuclear
constituents or/and their rearrangement at energies below $E_0$.

It is easy to check that
$$
\Psi(t)=U_1^{(1)}(t)+U_1^{(2)}(t)+\vphi_2\cdot u_2(t)
$$
where $U_1^{(1)}\in\cH_1$, $U_1^{(2)}\in Q_2\cH_2$ and
$u_2\in\C$ stand for components of solution
$U=\left(\begin{array}{c} U_1^{(1)}\\U_1^{(2)}\\ u_2
\end{array}\right)$, $U\in\tilde{\cH}$, of the evolutionary
problem

\begin{equation}
\label{FadEvol}
\begin{array}{l}
i\D\frac{dU}{dt}=H'_F U, \\
\reduction{U_1^{(1)}}{t=0}=\reduction{U_1^{(2)}}{t=0}=0, \,\,
\reduction{u_2}{t=0}=1
\end{array}
\end{equation}
corresponding to the Faddeev operator $H'_F$. Therefore,
to estimate the product
$$
\lal\Psi(t),\vphi_2\ral=\lal U_1^{(1)}(t),\vphi_2\ral +u_2(t),
$$
determining the probability $P_{\rm mol}(t)$, we use the
standard integral representation of function of operator via its
resolvent.  In the case considered we represent the evolution
operator for the problem~(\ref{FadEvol}) in terms of the
resolvent $(H'_F-z)^{-1}$,

\begin{equation}
\label{intexp}
\exp{\{-iH'_F t\}}=\D\frac{1}{2\pi i}\D\oint\limits_\gamma
dz \,{\rm e}^{-izt}(H'_F-z)^{-1}.
\end{equation}
Integration in~(\ref{intexp}) is provided along a contour
$\gamma$ going clockwise in the physical sheet around spectrum
of the Faddeev operator $H'_F$. Remember, this spectrum is real,
being a union of the Hamiltonians $h_0$ and $H$ spectra.  With
account of~(\ref{FadReduced}) and~(\ref{intexp}) one finds

\begin{equation}
\label{ProbScal}
\lal\Psi(t),\vphi_2\ral=\D\frac{1}{2\pi i}\oint\limits_\gamma
dz \, {\rm e}^{-izt}\,\D\frac{1-\lal g_1(z)b_{12},\vphi_2\ral}
{\lambda_2-z-\lal g_1(z)b_{12},b_{21}\ral}.
\end{equation}

In the physical sheet, norm of the generalized resolvent
$g_1(z)$ allows the estimate  $\|g_1(z)\|\leq{\mathop{\rm
dist}}^{-1}(z,\sigma(\tilde{H}_F))$ where  $\sigma(\tilde{H}_F)$
stands for spectrum of $\tilde{H}_F$.  Consequently, at $t=0$,
the integrand in~(\ref{ProbScal}) behaves like $z^{-1}$ as
$|\Img z|\to\infty$. One checks immediately due to this fact
that $\lal\Psi(0),\vphi_2\ral=1$.

Concerning properties of $\lal\Psi(t),\vphi_2\ral$ for $t>0$,
the following important statements take place.

{\rm 1.} Behavior of the integral~(\ref{ProbScal}) for $t>0$, in
the conditions of Section~\ref{Molecule}, is described by the
formula

\begin{equation}
\label{Prob0}
\lal\Psi(t),\vphi_2\ral=\exp\{-iz_{\rm mol} t\}+\varepsilon(t)
\end{equation}
where term $\varepsilon(t)=O(\|b_{12}\|)$ for all $t>0$
is small, $|\varepsilon(t)|\ll 1$.

{\rm 2.} If, additionally, the nuclear resonance $z_1$ is
``extremely'' narrow, i.~e.  $\Gamma_R^{(1)}\ll E_0-\lambda_2$,
then a more detailed representation takes place,

\begin{equation}
\label{ProbDetail}
\begin{array}{ccl}
\lal\Psi(t),\vphi_2\ral &=& \exp\{-iz_{\rm mol} t\}
\left[1-\D\frac{4a}{\left(\Gamma_R^{(1)}\right)^2} +
i\D\frac{2a_0}{\Gamma_R^{(1)}} +O(\|b_{12}\|)\right] + \\
 &+& \exp\{-iz_{\rm nucl} t\}
\left[\D\frac{4a}{\left(\Gamma_R^{(1)}\right)^2}-
i\D\frac{2a_0}{\Gamma_R^{(1)}} +O(\|b_{12}\|)\right] +
\tilde{\varepsilon}(t)
\end{array}
\end{equation}
with $a_0\equiv A\,\lal\vphi_1^{\rm res},\vphi_2\ral \lal
b_{12},\tvphi_1^{\rm res}\ral= A\,\lal\vphi_1^{\rm
res},\vphi_2\ral \lal v_1\vphi_2,\tvphi_1^{\rm res}\ral.$ As
in~(\ref{Prob0}), a background term
$\tilde{\varepsilon}(t)=O{(\|b_{12}\|)}$ is small,
$|\tilde{\varepsilon}(t)|\ll 1$, for all $t>0$.

Proof of the formula~(\ref{Prob0}) is carried out via estimating
a contribution to the integral~(\ref{ProbScal}) from the
resonance pole $z_{\rm mol}$ only. The
formula~(\ref{ProbDetail}) explicitly includes also a
contribution from the pole $z_{\rm nucl}$, because, with the
condition $\Gamma_R^{(1)}\ll E_0-\lambda_2$, one finds the
background summands are of a more high order of smallness as
compared to the contribution above from $z_{\rm nucl}$. The
estimation of the contributions from $z_{\rm mol}$ and $z_{\rm
nucl}$ is realized as a result of deforming the contour $\gamma$
fragments situated in a vicinity of the molecular energy
$\lambda_2$. A part of $\gamma$ situated initially in upper rim
of the cut, is pulled into the domain $\cD$ (see
Section~\ref{Molecule}) of a neighboring unphysical sheet.
Having done such a deformation one finds explicitly the residues
at $z=z_{\rm mol}$ and $z=z_{\rm nucl}$ of the integrand
in~(\ref{ProbScal}).  An analogous deformation of a part of
$\gamma$, situated initially in the lower rim, is realized in a
domain $\Img z<0$ of the physical sheet.  In such a way one
shows, for $t>0$, that the conjugate resonances
$\overline{z}_{\rm mol}$ and $\overline{z}_{\rm nucl}$ give to
the integral~(\ref{ProbScal}) only a very small contribution
which is then included into the background terms
$\varepsilon(t)$ and $\tilde{\varepsilon}(t)$.  At the same
time, the summands $\varepsilon(t)$ and $\tilde{\varepsilon}(t)$
include also a contribution to~(\ref{ProbScal}) from the
discrete spectrum  of $H$ (with ignorance of factors oscillating
when $t$ changes, a value of this contribution stays practically
the same for all $t$) as well as a contribution (decreasing
non-exponentially as $t\to\infty$) from the rest part of the
Hamiltonian $H$ continuous spectrum.

The formulae~(\ref{Prob0}) and~(\ref{ProbDetail}) show
explicitly, that, {\em in a large time interval $0\leq t<T$,
$T\sim \D\frac{2}{{\Gamma_R^{(m)}}} \bigl|\ln\mathop{\rm
max}|\varepsilon(t)|\bigr|$, decay of the ``molecular'' state
$\vphi_2$ in a presence of the narrow pre-threshold nuclear
resonance do has an exponential character%
\footnote{Surely, a similar statement takes place as well
in the case of the model Hamiltonians~(\ref{H3}) and~(\ref{H2}),
generally more simple than Hamiltonian~(\ref{Hmol}).
In particular, the two-channel model~(\ref{H2}) gives us for
$P_{\rm mol}(t)$ the following value:
$
P_{\rm mol}(t)=\left|\D\frac{1}{2\pi i}\oint_\gamma
dz \, {\rm e}^{-izt}\,
\biggl[\lambda_2-z-\lal r_1(z)b,b\ral\biggr]^{-1}\right|^2.
$
Here, the contour $\gamma$ goes clockwise around spectrum of the
operator~(\ref{H2}).  The function $P_{\rm mol}(t)$ represents a
probability for a time moment $t$ to find a system in the
initial ``molecular'' state described in the model~(\ref{H2}) as
$\vphi=\left(\begin{array}{c} 0\\1 \end{array} \right)$.  The
integral included in the expression above for $P_{\rm mol}(t)$,
is quite analogous to~(\ref{ProbScal}).  One can show in
conditions of Section~\ref{Two-Channel} that all the dependence
of the integral on $t$ is described exactly by the
formula~(\ref{Prob0}) or by a more detail formula
like~(\ref{ProbDetail}). }.
A rate of this decay is determined mainly by the width
$\Gamma_R^{(m)}$ of the ``molecular'' resonance $z_{\rm mol}$,
i.~e. by a concrete value of the ratio $|\Real
a|/\Gamma_R^{(1)}$:}

\begin{equation}
\label{Exponent}
P_{\rm mol}(t)\cong\exp\{-{\Gamma_R^{(m)}} t\}
\cong\exp\left\{-\frac{4|\Real a|}{\Gamma_R^{(1)}}\,t\right\}.
\end{equation}

\acknowledgments
The authors would like to thank Professor~A.~M.~Baldin for interest 
to this research and support.
  
The work was partly supported by the Scientific Division of
NATO, \hbox{Project Nr.~930102,} and the Russian Foundation for
Basic Research, Projects \hbox{Nr.~96-01-01292}, 
\hbox{Nr.~96-01-01716} and
\hbox{Nr.~96-02-17021.}

\end{document}